\documentclass[12pt]{iopart}

\usepackage [dvips]{graphicx}
\usepackage{amssymb}
\usepackage{epsfig}

\usepackage{graphicx}
\usepackage{verbatim}

\usepackage[T1]{fontenc}

\begin{document}

\newcommand{\beq}{\begin{equation}}
\newcommand{\eeq}{\end{equation}}
\newcommand{\bea}{\begin{eqnarray}}
\newcommand{\eea}{\end{eqnarray}}
\newcommand{\si}{\sigma_i^z}
\newcommand{\sj}{\sigma_j^z}


\title[Condensation of the roots of real random polynomials on the real axis]
{Condensation of the roots of real random polynomials on the real axis}

\author{Gr\'egory Schehr$^1$, Satya N. Majumdar$^2$}

\address{$^1$ Laboratoire de Physique Th\'eorique, Universit\'e de
  Paris-Sud,  91405 Orsay France \\  
$^2$ Laboratoire de Physique Th\'eorique et Mod\`eles
  Statistiques, Universit\'e Paris-Sud, B\^at. 100, 91405 Orsay Cedex,
France}

\begin{abstract}
We introduce a family of real random polynomials of degree $n$ whose
coefficients $a_k$  
are symmetric independent Gaussian variables with variance $\langle
a_k^2\rangle~=~e^{-k^\alpha}$, indexed by a real $\alpha \geq 0$. We compute
exactly the mean number of real roots $\langle N_n \rangle$ for large $n$. As
$\alpha$ is varied, one finds three different phases. First, for $0
\leq \alpha < 1$, one finds that $\langle N_n \rangle \sim
(\frac{2}{\pi}) \log{n}$. For $1 <
\alpha < 2 $, there is an intermediate phase where $\langle N_n
\rangle$ grows algebraically with a continuously varying exponent, 
$\langle N_n \rangle \sim \frac{2}{\pi}
\sqrt{\frac{\alpha-1}{\alpha}} \, n^{\alpha/2}$. And finally for
$\alpha > 2$, one finds a third phase where $\langle N_n \rangle \sim
n$. This family of real random
polynomials thus exhibits a condensation of their roots on the real
line in the sense that, for large $n$, a finite fraction of their roots $\langle N_n\rangle/n$ are real. 
This condensation occurs via a localization of the real roots
around the values $\pm
\exp{\left[\frac{\alpha}{2}(k+\frac{1}{2})^{\alpha-1} \right]}$, $ 1 \ll k \leq n$. 
\end{abstract}

\maketitle

\section{Introduction}

Since the early work of Bloch and P{\'o}lya \cite{bloch} in the 30's, the study of random algebraic equations has now a long story \cite{bharucha, farahmand}. In the last few years, it attracted a renewed interest in the context of probability and number theory \cite{edelman}, as well as in the field of quantum chaos
\cite{bogo}. Recently, we showed that there are also interesting
connections between random polynomials and persistence properties of
physical systems \cite{us_short, us_long}.  

Here we consider real random polynomials, {\it i.e.} polynomials with
real random coefficients, of degree $n$. While these polynomials have exactly $n$ roots in the complex plane, 
the number of roots on the {\it real} line $N_n$ is a random variable. One would like to characterize the
statistics of this random variable and a natural question is thus :
what is the mean number $\langle N_n \rangle$ of real roots and how
does it behave with $n$ for large $n$ \cite{edelman}? This question 
has been widely studied in the past for Kac's polynomials $K_n(x) =
\sum_{k=0}^n a_k\, x^k$ where $a_k$ are independent and identically
distributed (i.i.d.) random variables of
finite variance $\langle a_k^2 \rangle = \sigma^2$. In that case it is
well known 
that $\langle N_n \rangle \sim \frac{2}{\pi}\log 
n$, independently of $\sigma$. This result was first obtained by
Kac \cite{kac} for Gaussian 
random variables and it was later shown to hold also for a wider class
of distributions of the coefficients $a_k$ \cite{bharucha,
  farahmand}. Interesting generalizations of Kac's polynomials have
been studied in the literature where $a_k$ are independent Gaussian variables but
non identical, such that  
$\langle a_k^2\rangle = k^{d-2}$, where $d>0$ is a real number,
leading to $\langle N_n \rangle  \sim \pi^{-1}(1+\sqrt{d/2})\log{n}$
\cite{us_long, das}. Given the robustness of this asymptotic
logarithmic behavior of $\langle N_n \rangle$, it is natural to search for random
polynomials for which $\langle N_n \rangle$ increases faster than $\log{n}$, for instance algebraically. 

One such instance is provided  by the real Weyl polynomials $W_n(x)$
defined by  
\begin{eqnarray}\label{weyl}
W_n(x) = \sum_{k=0}^n \epsilon_k \frac{x^k}{\sqrt{k!}} \;,
\end{eqnarray}
where $\epsilon_k$ are i.i.d. random variables of zero mean and unit
variance. Thus here, $a_k = \epsilon_k/\sqrt{k!}$ and the variance is
$\langle a_k^2 \rangle = 1/k!$, which for large $k$ behaves as
$\langle a_k^2 \rangle \propto e^{-k \log k}$. For these real polynomials
in Eq. (\ref{weyl}), it is known that $\langle N_n \rangle \propto
n^{1/2}$. For instance, in the special case where $\epsilon_k$ are
Gaussian random  
variables of unit variance, one has $\langle N_n \rangle \sim
\frac{2}{\pi} \sqrt{n}$ \cite{us_long, leboeuf}. Another interesting
and intriguing instance of real random polynomials was introduced a
long time ago by Littlewood and Offord \cite{littlewood} who studied
the random polynomials $L_n(x)$ given by 
\begin{eqnarray}\label{little}
L_n(x) = \frac{1}{2}+\sum_{k=1}^n \epsilon_k \frac{x^k}{(k!)^{k}} \;,
\end{eqnarray}
where $\epsilon_k = \pm 1$ with equal probability. Thus in this case
$a_k = \epsilon_k /(k!)^{k}$ and the variance is $\langle a_k^2
\rangle = 1/(k!)^{2k}$, which behaves for large $k$ as $\langle a_k^2
\rangle \propto e^{-2k^2 \log k}$. Using algebraic
methods, they showed that such polynomials $L_n(x)$ have all their
roots real and therefore $\langle N_n \rangle = n$.   

We thus have here two examples of real random polynomials in
Eq.~(\ref{weyl}) and Eq.~(\ref{little}) where, at variance with Kac's
polynomials, $\langle N_n \rangle$ grows algebraically with $n$. In the
second example (\ref{little}), the number of real roots is
``macroscopic'' in the sense that, for large $n$, there is a finite
fraction $\langle 
N_n \rangle/n$ of the roots which are on the real axis. For $L_n(x)$
in Eq. (\ref{little}) this fraction is exactly one. We thus say
that there is a {\it condensation} of the roots on the real line,
similar to a Bose-Einstein condensation where a finite fraction of the
particles of a quantum-mechanical system (Bosons) condense into the lowest
energy level. In the case of random polynomials, the roots play the
role of the particles and the equivalent of the ground state is the real line.

\begin{figure}
\includegraphics[angle=0,scale=0.6]{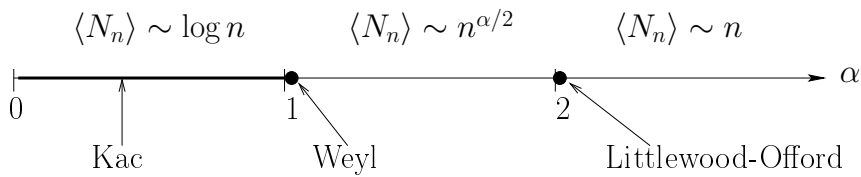}
\caption{Asymptotic behavior of the mean number of real roots $\langle
  N_n \rangle$ of $P_n(x)$ in Eq. (\ref{def_poly}) as a function of
  $\alpha$. These polynomials exhibit a condensation of their roots on
  the real axis for $\alpha \geq 2$.}\label{fig1}
\end{figure}

The purpose of this paper is to understand what types of polynomials
lead to this condensation phenomenon. Of course, it is very difficult to
address this question for any random coefficients $a_k$. However,
guided by the two examples above in Eq.~(\ref{weyl}) and
Eq.~(\ref{little}), and in particular by the large $k$ behavior of 
$\langle a_k^2 \rangle$, we introduce a family of random polynomials
$P_n(x)$ indexed by a real $\alpha \geq 0$ defined by  
\begin{eqnarray}\label{def_poly} 
P_n(x) = \sum_{k=0}^{n} a_k \, x^k \;, \; \langle a_k^2 \rangle = e^{-k^\alpha} \;,
\end{eqnarray} 
where $a_k$ are real independent Gaussian random variables of zero
mean. While $\alpha=0$ corresponds to Kac's polynomials, we recall that, for $W_n(x)$ in Eq. (\ref{weyl}), $\langle
a_k^2 \rangle \propto e^{-k \log k}$ and for $L_n(x)$ in
Eq. (\ref{little}), $\langle a_k^2 \rangle \propto e^{-2 k^2 \log
  k}$. Therefore, due to the extra 
logarithmic factor, these random polynomials are not exactly of the form
introduced above (\ref{def_poly}). However, for $\alpha \to 1^+$, one
expects to recover the behavior of $W_n(x)$ in Eq. (\ref{weyl}) while
for $\alpha \to 2^+$, one
expects $P_n(x)$ to behave similarly to $L_n(x)$ in
Eq. (\ref{little}) : this is depicted schematically in Fig. \ref{fig1}. 

Our main results can be summarized as follows. As $\alpha \geq 0$ is varied
one finds three different {\it phases}. The first phase corresponds to $0 \leq
\alpha < 1 $, where one finds that $\langle N_n \rangle \sim (2/\pi) \log{n}$. In the second one,
corresponding to $1 < \alpha < 2$, one has  $\langle N_n \rangle \sim 
\frac{2}{\pi}\sqrt{\frac{\alpha-1}{\alpha}} \, n^{\alpha/2}$. And in the third phase, for $\alpha > 2$, one
finds $\langle N_n \rangle \sim n$. The condensation of the roots
on the real axis thus happens for $\alpha \geq 2$ and as one increases
$\alpha$, the condensation transition sets in at the critical value
$\alpha_c = 2$. Furthermore, one finds that these real
roots condense into a quasi-periodic structure such that there is, on
average, one root in the interval  
$[-x_{m+1},-x_m] \cup [x_m,x_{m+1}]$, with $x_m =
e^{\frac{\alpha}{2}m^{\alpha-1}}$, with $1 \ll m <n$. These different behaviors
are summarized in Fig.~\ref{fig1}. By analogy with phase transitions
of statistical systems the case $0 < \alpha < 1 $ can be
considered as a high-temperature phase whereas $\alpha > 2$
corresponds to the low-temperature (ordered) phase.     
Roughly speaking, one can consider our results as an interesting example where the transition from the
high temperature where $\langle N_n \rangle \propto \log{n}$ (governed
by a ``$\alpha = 0$ fixed point'') to the
low temperature phase where $\langle N_n \rangle \propto n$ (governed
by ``$\alpha = \infty$'' fixed point) happens through a  
{\it marginal phase}, for $1< \alpha < 2 $, where $\langle N_n \rangle
\sim n^{\phi}$ with an exponent $\phi = \alpha/2$ which depends
continuously on $\alpha$. 

The paper is organized as follows. In section 2, we describe the general
framework to compute the local density of real roots, which directly
leads to $\langle N_n \rangle$. In section 3 to 6 we then analyse
separately the 
cases $0 \leq \alpha < 1$, $\alpha < 2$, $\alpha > 2$ and the
''critical case'' $\alpha =  2$. In section 7, we
give a qualitative argument to explain the condensation transition
occurring at 
$\alpha_c = 2$ before we conclude in section 8. The Appendix contains
some useful technical details.   
 
\section{General framework}

First we notice that given that $P_n(x)$, as a function of
$x$, is a Gaussian process, it is completely characterized by its
two-point correlation function $C_n(x,y)$
\beq\label{def_correl}
C_n(x,y) = \langle P_n(x) P_n(y) \rangle = \sum_{k=0}^n e^{-k^\alpha} \,x^k\,y^k \;,
\eeq
where we used the notation $\langle ... \rangle$ to denote an average over the random variables $a_k$. 
A central object involved in the calculation of  $\langle N_n \rangle$
is $\rho_n(x)$, the mean density of real roots at point $x$. If we denote $\lambda_1, \lambda_2, ..., \lambda_p$ the $p$ real
roots (if any) of $P_n(x)$, one has $\delta(P_n(x)) = \sum_{i=1}^p \delta(x-\lambda_i)/|P_n'(\lambda_i)|$ such that 
$\rho_n(x)$ can be written as
\bea
\rho_n(x) &=& \sum_{i=1}^p \langle \delta(x-\lambda_i) \rangle = 
\langle |P_n'(x)|\delta(P_n(x)) \nonumber \\
&=& \int_{-\infty}^\infty dy |y| \langle \delta(P_n'(x)-y) \delta(P_n(x)) \rangle  \;. \label{def_density}
\eea
Under this form (\ref{def_density}), one observes that the computation of 
the mean density involves the joint distribution of the polynomial
$P_n(x)$ and its derivative $P'_n(x)$ which is simply a bivariate
Gaussian distribution. After Gaussian integration over $y$, one obtains 
\begin{eqnarray}\label{def_density_inter}
&&\rho_n(x) = \frac{\sqrt{c_n(x) (c_n'(x)/x + c_n''(x)) - [c_n'(x)]^2
}}{2 \pi c_n(x)} \;, \\
&&c_n(x) = C_n(x,x) = \sum_{k=0}^n e^{-k^\alpha} x^{2k}\;. \nonumber
\end{eqnarray}
This formula (\ref{def_density_inter}) can be written in a very
compact way \cite{edelman} : 
\begin{eqnarray}
\rho_n(x) = \frac{1}{\pi} \sqrt{\partial_u \partial_v \log C_n(u,v)}
\bigg 
|_{u=v=x} \;.\label{ek_formula}
\end{eqnarray}   
Given that the random coefficients $a_k$ are drawn from a symmetric distribution, we can restrict our study of $\rho_n(x)$ on ${\mathbb {R}}^+$ from which one obtains the mean number of real roots $\langle N_n \rangle$ as
\beq
\langle N_n \rangle = 2 \int_0^\infty \rho_n(x) dx \;.
\eeq

{\bf An important change of variable.} We will see below that it is useful to consider these polynomials $P_n(x)$ in terms of another variable
$Y$ defined as
\beq\label{new_variable}
Y = \left(\frac{2}{\alpha} \log{x} \right)^{\frac{1}{\alpha-1}} \;.
\eeq
We denote $\hat \rho_n(Y)$ the mean density of the real roots in terms of this new variable such that one 
has also $\langle N_n \rangle = \int_0^\infty \hat \rho_n(Y) dY$. For $0 < \alpha < 1$ we will see that, for large $n$, most of the real roots of $P_n$ are located close to $Y = n$ while for $\alpha > 1$, the density extends over the whole interval $Y \in [1,n]$. This change of variable (\ref{new_variable}) is motivated by the following analysis. 

First we notice that $C_n(x,y) = \sum_{k=0}^n e^{-k^\alpha} x^k y^k$ in Eq. (\ref{def_correl}) is of the form $C_n(x,y)=
c_n(\sqrt{x y})$. Anticipating a saddle point analysis, one writes $c_n(x)$ as 
\bea\label{def_series}
&&c_n(x) = \sum_{k=0}^n e^{-k^\alpha} x^{2k} = \sum_{k=0}^n
\exp{\left(-\phi(k,x)   \right)} \;, \;\phi(k,x) = k^\alpha - 2 k
\log{x} \;. 
\eea
Although $\phi(k,x)$ is defined for integers $k = 0, 1, 2, \cdots, n$,
it is readily extended to the real axis and denoted $\phi(u,x) =
u^\alpha - 2u\log{x}$ for $u \in
\mathbb{R}^+$. The behavior of $c_n(x)$ is essentially governed by the
behavior of $\phi(u,x)$ as a function of $u$ (and fixed $x$). In
particular, for $\alpha < 1$, $\phi(u,x)$ has a single maximum while
for $\alpha > 1$, it has a single minimum for $u = u^*(x)$ given by
\bea\label{def_ustar} 
&&\partial_u \phi(u^*(x),x) = 0 \; , \;  \partial^2_u \phi(u^*(x),x) =
\alpha(\alpha-1) u^*(x)^{\alpha-2} > 0 \;, \nonumber \\
&& u^*(x) = \left(\frac{2}{\alpha} \log{x} \right)^{\frac{1}{\alpha-1}} \;.
\eea 
%
The new variable $Y$ introduced above in Eq. (\ref{new_variable}) is
thus precisely $Y = u^*(x)$. As a consequence, the density behaves
quite differently in both cases $\alpha < 1$ and $\alpha > 1$.

For $\alpha < 1$, most of the real roots on $\mathbb{R}^+$ are located
in $[1, \infty]$. For 
fixed $x>1$, $\phi(u,x)$ as a function of $u$ in the interval $[0,n]$ has a
global minimum for $u=n$. Therefore, the sum entering in the expression of 
$c_n(x)$ in Eq. (\ref{def_series}) will be dominated by the terms with $k \sim n$. The expansion
of $\phi(k,x)$ in Taylor series around $k=n$ yields
\bea
\phi(k,x) &=& \phi(n,x) + (k-n) (\alpha n^{\alpha-1} - 2 \log{x} ) + \cdots \nonumber \\
&=& (1-\alpha)n^{\alpha} - k(\alpha n^{\alpha-1}-2\log{x}) + \cdots \;,
\eea
where the higher order terms can be neglected in the large $n$ limit because $\partial^j \phi(n,x)/\partial u^j = {\cal O}(n^{\alpha - j})$ for $j \geq 2$. Thus, for $\alpha < 1$ one has 
\bea\label{kac_sim}
c_n(x) \sim e^{-(1-\alpha)n^\alpha} \sum_{k=0}^n (x e^{-\frac{\alpha}{2} n^{\alpha-1}})^{2k} \;,
\eea
which, in terms of the rescaled variable $\tilde x = x \,e^{-\frac{\alpha}{2} n^{\alpha-1}}$, is the correlator of Kac's polynomials. From this observation (\ref{kac_sim}), one can straightforwardly obtain the mean number of real roots $\langle N_n \rangle$, this will be done in section 3. 

For $\alpha > 1$, the situation is quite different and in that case, $\phi(u,x)$ has a single
minimum for $u = u^*(x) = (\frac{2}{\alpha}
\log{x})^{\frac{1}{\alpha-1}}$ (\ref{def_ustar}). Besides, we will see 
below that the main contribution 
to $\langle N_n \rangle$ on  $\mathbb{R}^+$ comes from the interval $1 < x < \exp{\left(\frac{\alpha}{2} n^{\alpha-1} \right)}$ where $1<
u^*(x) < n$. In that case the sum entering in the definition of
$c_n(x)$ in Eq. (\ref{def_series}) is indeed dominated by $k \sim
u^*(x)$ and $c_n(x)$ can be evaluated by a saddle point
calculation. For this purpose, one obtains after some algebra explained in the Appendix, a convenient expression of $\rho_n(x)$ as
\bea\label{start_expr_rho}
\hspace*{-1cm}\rho_n(x) = \frac{1}{\pi x} \left(  \frac{ \sum_{k=0}^n (k-u^*(x))^2  e^{-\phi(k,x)}}{\sum_{k=0}^n e^{-\phi(k,x)}} - 
 \left[ \frac{ \sum_{k=0}^n (k-u^*(x))  e^{-\phi(k,x)}}{\sum_{k=0}^n e^{-\phi(k,x)}} \right]^2
\right)^{\frac{1}{2}} \;,
\eea
which is the starting point of our analysis for $\alpha > 1$. For $1 < x < \exp{\left(\frac{\alpha}{2} n^{\alpha-1} \right)}$, one has $u^*(x)<n$ so that the sums over $k$ in Eq. (\ref{start_expr_rho}) are dominated
by $k \sim u^*(x)$. The Taylor expansion of $\phi(k,x)$ around this minimum reads
\beq\label{taylor_exp}
\phi(k,x) = \phi(u^{*}(x),x) + \sum_{j=2}^\infty \frac{\alpha (\alpha-1)...(\alpha-j+1)}{j !} (k-u^*(x))^j [u^*(x)]^{{\alpha-j}} \;.
\eeq
For large $x$, $u^*(x) \propto (\log{x})^{1/(\alpha-1)}$ is also large so that, to leading order in $x$,  
one can retain only the term corresponding to $j=2$ in the Taylor expansion in Eq. (\ref{taylor_exp}). This yields, for large $x$
\bea\label{inter}
&&\sum_{k=0}^n g(k-u^*(x)) \exp{\left(-\phi(k,x) \right)}  \\
&& \sim e^{-\phi(u^*(x),x)}  \sum_{k=0}^n g(k-u^*(x))
\exp{\left[-\frac{\alpha(\alpha-1)}{2}(k-u^*(x)) [u^*(x)]^{\alpha-2} \right]}
\;, \nonumber
\eea
with $g(z) = z$ or $g(z) = z^2$ as in Eq. (\ref{start_expr_rho}). For later purpose it is useful to write $u^*(x) = \lfloor u^*(x)\rfloor + b$ 
 with $0 < b < 1$, where $\lfloor u^*(x) \rfloor$ is the largest integer smaller than $u^*(x)$ ({\it i.e.} the floor function). Performing the change of variable
$m = k - \lfloor u^*(x)\rfloor$ in the discrete sum (\ref{inter}), such that $k - u^*(x) = m - b$ one obtains the useful expression 
\bea
&&\sum_{k=0}^n
  g\left( k-u^*(x)\right) \exp{(-\phi(k,x))}  \\
&&\sim e^{-\phi(u^*(x),x)} \sum_{m = -\lfloor u^*(x) \rfloor}^{n-\lfloor u^*(x) \rfloor} g(m -
  b)  \exp{\left[-\frac{\alpha
	(\alpha-1)}{2} 
    (m-b)^2 [u^*(x)]^{{\alpha-2}}\right]} \nonumber \;.
  \label{asympt_largex} 
\eea
One clearly sees in expression (\ref{asympt_largex}) that the
behavior of this discrete sum, due to the term $[u^*(x)]^{\alpha-2} \propto (\log{x})^{(\alpha-2)/(\alpha-1)}$, will depend on the sign of
$\alpha-2$. We will thus treat the three cases $1 < \alpha < 2$,
$\alpha>2$ and $\alpha=2$ separately. This will be done in section 4, 5 and 6 respectively.

\section{The case $0 < \alpha < 1$}

In that case, from the expression for $c_n(x)$ in Eq. (\ref{kac_sim}),
we can use the results of Kac's polynomials to obtain that most of the
real roots will be such that, for large $n$, $x
e^{-\frac{\alpha}{2}n^{\alpha-1}} -1 = {\cal O}(n^{-1})$
\cite{fyodorov}.  
In other words, the real roots are distributed in a region of width
$1/n$ around $e^{\frac{\alpha}{2}n^{\alpha-1}} = 1 + \frac{\alpha}{2}
n^{\alpha-1} + {\cal O}(n^{\alpha-2})$ and this distribution is
exactly the same as the one for Kac's polynomials (corresponding to
$\alpha=0$). The number of real roots is thus also the same and given
by 
\beq\label{last_kac}
\langle N_n \rangle \sim \frac{2}{\pi} \log{n} \;,
\eeq
independently of $\alpha < 1$.

\section{The case $1 < \alpha < 2$}

In that case $[u^*(x)]^{\alpha-2} \to 0$ for large $u^*(x)$ and one thus sees 
on the asymptotic expression in
Eq. (\ref{asympt_largex}) that the discrete sum can be replaced by an
integral.  This yields, for large $n$ and large $x$ with $x <
\exp{(\frac{\alpha}{2} n^{\alpha-1}})$ 
\beq\label{discrete_integral}
\hspace*{-0.5cm}\sum_{k=0}^n
  g\left( k-u^*(x)\right) \exp{(- \phi(k,x))} \sim e^{-\phi(u^*(x),x)}\int_{-\infty}^\infty  g(y)
  e^{-\frac{\alpha(\alpha-1)}{2} y^2 u^*(x)^{{\alpha-2}} } \, dy \;.
\eeq
Note that the prefactor $e^{-\phi(u^*(x),x)}$ is unimportant for the computation of $\rho_n(x)$ because it disappears between the numerator and the denominator in Eq. (\ref{start_expr_rho}) and it will be omitted below. In particular, setting $g(z) = 1$ in Eq. (\ref{discrete_integral}) one has
\bea\label{eq_g1}
\sum_{k=0}^n  \exp{(-\phi(k,x))} \propto \sqrt{2 \pi}
\left[\frac{u^*(x)^{2-\alpha}}{\alpha(\alpha-1)} \right]^{\frac{1}{2}} \;,
\eea
and similarly, setting $g(z)=z^2$ in Eq. (\ref{discrete_integral}) one has
\bea\label{eq_gx2}
\sum_{k=0}^n  (k-u^*(x))^2 \exp{(-\phi(k,x))} \propto \sqrt{2 \pi}
\left[\frac{u^*(x)^{2-\alpha}}{\alpha(\alpha-1)} \right]^{\frac{3}{2}} \;,
\eea
while $\sum_{k=0}^n  (k-u^*(x))\exp{(-\phi(k,x))} \sim 0$ to
lowest order in $n$.  Therefore using the exact expression given in
Eq. (\ref{start_expr_rho}) together with the 
asymptotic behaviors given in Eq.~(\ref{eq_g1}, \ref{eq_gx2}), one obtains
the large $x$ behavior of 
$\rho_n(x)$ as
\beq\label{asympt_largex_alleq2}
\rho_n(x) \sim  \frac{1}{\pi x}
\frac{1}{\sqrt{\alpha(\alpha-1)}} \left(\frac{2}{\alpha} \log{x}
\right)^{\frac{2-\alpha}{2(\alpha-1)}} \;.
\eeq 
For a clear comparison with the case $\alpha > 2$ (which will be analysed in
the next section), it is convenient to
write the density $\hat \rho_n(Y)$, in terms of the variable $Y =
\left(\frac{2}{\alpha} \log{x} \right)^{\frac{1}{\alpha-1}}$, 
which reads, for $1 \ll Y < n$ 
\bea\label{asympt_largeX_alleq2}
\hat \rho_n(Y) \sim \frac{\sqrt{\alpha(\alpha-1)}}{2 \pi} Y^{-\frac{1}{2}(2-\alpha)} \;,
\eea
and in Fig. \ref{fig2} a), we show a sketch of this asymptotic
behavior (\ref{asympt_largeX_alleq2}) of $\hat \rho_n(Y)$ for $1 \ll Y
< n$. 

We can now compute $\langle N_n \rangle = \int_{-\infty}^\infty  \rho_n(x) \, dx$. First, one notices that for $\alpha > 1$, the series entering in the definition of $c_n(x)$
in Eq. (\ref{def_series}) has an infinite radius of convergence so
that one readily obtains that $\int_{-1}^{+1} \rho_n(x) \, dx$ is of
order ${\cal O}(1)$ in the 
limit $n \to \infty$. Besides, for large $x \gg e^{\frac{\alpha}{2}
  n^{\alpha-1}}$, one has (see also Ref. \cite{us_long}) 
\beq\label{very_largeX}
\rho_n(x) \sim \sqrt{\frac{\langle a_{n-1}^2\rangle}{\langle a_{n}^2
    \rangle}}\frac{1}{\pi x^2} \sim \frac{e^{\frac{\alpha}{2}
    n^{\alpha-1}}}{\pi x^2} \;,
\eeq 
which implies that $\int_{e^{\frac{\alpha}{2}
    n^{\alpha-1}}}^\infty  \rho_n(x) \, dx$ is also of order ${\cal
  O}(1)$ in the limit $n\to \infty$. From these properties, it follows
that the main contributions to $\langle N_n \rangle$ on ${\mathbb R}^+$ comes from the
interval $[1, e^{\frac{\alpha}{2} n^{\alpha-1}}]$ where the asymptotic
behavior of $\rho_n(x)$ is given in
Eq. (\ref{asympt_largex_alleq2}). Therefore one has
\beq\label{N_alleq2}
\langle N_n \rangle \sim 2 \int_1^{e^{\frac{\alpha}{2} n^{\alpha-1}}} \rho_n(x) \, dx
\sim \frac{2}{\pi} \sqrt{\frac{\alpha-1}{\alpha}} \,n^{\alpha/2} \;,
\eeq
where the factor $2$ comes from the additional contribution coming from $[-e^{\frac{\alpha}{2} n^{\alpha-1}},-1]$. We thus have here an algebraic growth
$\langle N_n \rangle \propto n^{\alpha/2}$ with a continuously varying exponent $\alpha/2$. This exponent tends to $1/2$ as $\alpha \to 1^+$, which is
expected from the analysis of Weyl polynomials $W_n(x)$ in Eq. (\ref{weyl}) for which $\langle a_k^2 \rangle \propto e^{-k \log k}$ (although the variance is not exactly of the form $\langle a_k^2 \rangle = e^{-k^\alpha}$). Besides, from Eq. (\ref{N_alleq2}), one also obtains that 
the amplitude of this term proportional to $n^{\alpha/2}$ vanishes when $\alpha \to 1$. We recall that for $\alpha \leq 1$, one has instead $\langle N_n \rangle \propto (\frac{2}{\pi}) \log{n}$ (\ref{last_kac}), characteristic for Kac's polynomials. This suggests that this limit $\alpha \to 1$ is rather singular in the sense that the asymptotic behavior of $\langle N_n \rangle$
for large $n$ changes "discontinuously"  from $\log{n}$ to $\sqrt{n}$.

\section{The case $\alpha > 2$}

In that case, the behavior of the discrete sum in
Eq. (\ref{asympt_largex}), which 
enters in the computation of $\rho_n(x)$ (\ref{start_expr_rho}) is quite
different. Indeed, in that case $[u^*(x)]^{\alpha-2} \propto (\log{x})^{(\alpha-2)/(\alpha-1)}\to \infty$ for large
$x$ and therefore the leading term for large $x$ in
Eq.~(\ref{asympt_largex}) corresponds to $m=0$ if $b < 1/2$ or $m=1$ in
$b>1/2$.  
Keeping these leading contributions, one has
\bea
&&\sum_{k=0}^n
  g\left( k-u^*(x)\right) \exp{(-\phi(k,x))} \propto g(-b) \exp{\left[-\frac{\alpha (\alpha-1)}{2} b^2 u^*(x)^{{\alpha-2}}\right]} \nonumber \\
&&+ g(1-b)\exp{\left[-\frac{\alpha (\alpha-1)}{2} 
    (1-b)^2 u^*(x)^{{\alpha-2}}\right]}\label{asympt_largex_alphaleq2} \;.
\eea
where, again, we have omitted the unimportant prefactor $e^{-\phi(u^*(x),x)}$. Using this large $x$ expansion (\ref{asympt_largex_alphaleq2}), one obtains $\rho_n(x)$ in Eq. (\ref{start_expr_rho}) as
\bea
  \rho_n(x) \sim \frac{2}{(\pi x)\cosh{\left[\frac{\alpha(\alpha-1)}{2} Y^{\alpha-2}(1-2b)  \right]}   } \, , \, Y =  \left(\frac{2}{\alpha} \log{x} \right)^{\frac{1}{\alpha-1}} \;.
\eea
In terms of the variable $Y$, the density $\hat \rho_n(Y)$ reads, 
\bea\label{pseudo_periodic}
\hat \rho_n(Y = \lfloor Y \rfloor + b) \sim \frac{\alpha(\alpha-1)
  Y^{\alpha-2}}{2\pi\cosh{\left[  \frac{\alpha(\alpha-1)}{2}
      Y^{\alpha-2}(1-2b)   \right]}} \;. 
\eea
In Fig. \ref{fig2} c), one shows a sketch of $\hat \rho_n(Y)$ for
large $Y < n$ given by Eq. (\ref{pseudo_periodic}) : it is
qualitatively very different from the case
$\alpha < 2$ (see Fig. \ref{fig2} a)). Indeed, $\hat \rho_n(Y)$
exhibits peaks centered around $k + \frac{1}{2}$ for large integers
$1 \ll k < n$. The height of these peaks is given by $\alpha(\alpha-1)
k^{\alpha-2}/(2 \pi)$ whereas its width scales like $k^{2 - \alpha}$.   

From $\rho_n(x)$, one can now compute the mean number of real
roots. As in the case $\alpha < 2$ (see Eq. (\ref{very_largeX}) and above), one
can show that the main contribution to $\langle N_n \rangle$ comes
from the intervals $[-e^{\frac{\alpha}{2} n^{\alpha-1}},-1]$ and 
$[1, e^{\frac{\alpha}{2} n^{\alpha-1}}]$. One thus
has from Eq. (\ref{pseudo_periodic})
\bea\label{condensation}
\langle N_n \rangle &=& 2 \int_0^\infty \rho_n(x) \, dx \sim 2 \int_0^n  \hat
\rho_n(Y) \, dY \\
&\sim&
 \sum_{k \gg 1}^n \int_0^1 \frac{\alpha(\alpha-1)
   k^{\alpha-2}}{\pi\cosh{\left[  \frac{\alpha(\alpha-1)}{2}
       k^{\alpha-2}(1-2b)   \right]}} \,db  \sim \sum_{k \gg 1}^n
 \int_{-\frac{\alpha(\alpha-1)}{2}k^{\alpha-2}}^{\frac{\alpha(\alpha-1)}{2}k^{\alpha-2}} \frac{dz}{\pi \cosh{z}} \,, \nonumber
\eea
and finally 
\bea
\langle N_n \rangle \sim n \;,
\eea
where we have used $\int_{-\infty}^\infty dz/\cosh{z} = \pi$. This
condensation of the roots on the real axis, characterized by the fact
that $\langle N_n \rangle \sim n$ thus occurs via the
formation of this quasi-periodic structure (see Fig. \ref{fig2}
c)). More precisely, this computation in Eq. (\ref{condensation})
shows that for large $k$, $2 \int_k^{k+1} \hat \rho_n(Y) \,dY \sim 1$ which
means, going back to the original variable $x$, that there is, on
average, one root in the interval $[-x_{k+1},-x_k] \cup [x_k,x_{k+1}]$,
with $x_k = e^{\frac{\alpha}{2} k^{\alpha-1}}$.

\section{The special case $\alpha = 2$}

\begin{figure}
\includegraphics[angle=0,scale=0.8]{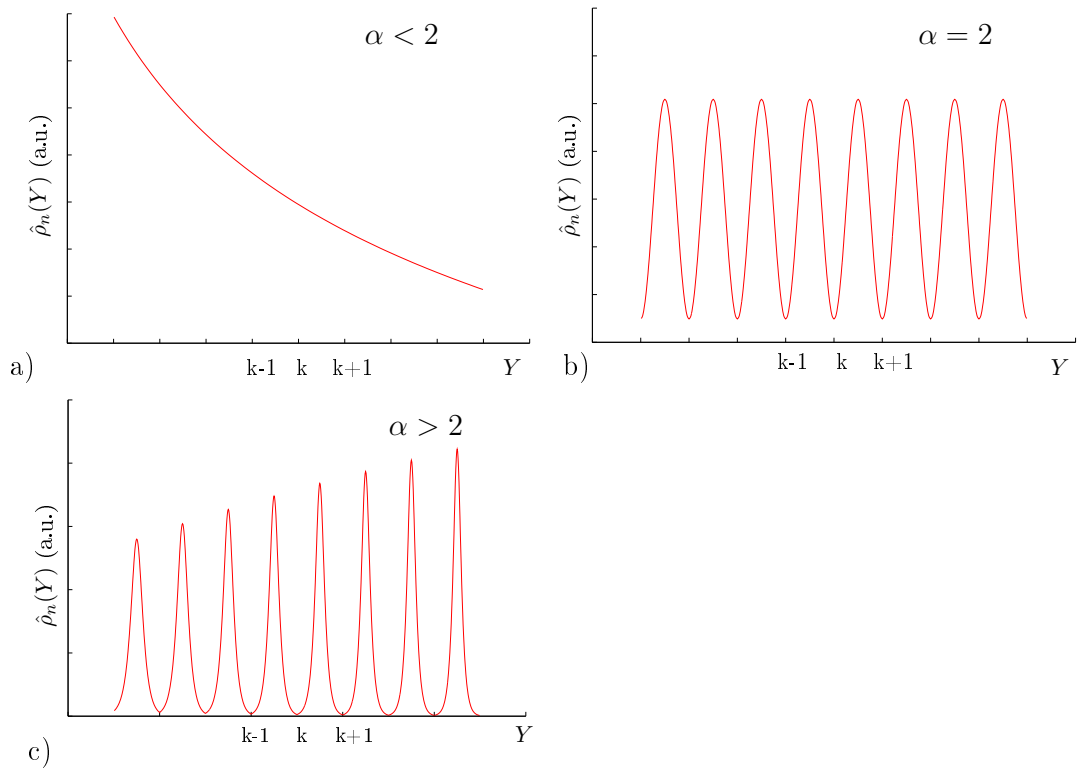}
\caption{{\bf a)} : Sketch of $\hat \rho_n(Y)$ (in arbitrary units)
  given in Eq. (\ref{asympt_largeX_alleq2}) as 
  a function of $Y$ for $1 \ll Y < n$ for $\alpha < 2$. {\bf b)} : Sketch of $\hat \rho_n(Y)$ (in arbitrary units)
  given in Eq. (\ref{start_expr_rho_mu}) as 
  a function of $Y$ for $1 \ll Y < n$ for $\alpha = 2$. {\bf c)} : Sketch of $\hat \rho_n(Y)$ (in arbitrary units)
  given in Eq. (\ref{pseudo_periodic}) as 
  a function of $Y$ for $1 \ll Y < n$ for $\alpha > 2$. Here $k$
  denotes an integer with $1 \ll k < n$.}\label{fig2} 
\end{figure}

In view of the previous analysis, it is tempting to consider the fraction of real roots $\Phi = \lim_{n
  \to \infty} \langle N_n \rangle / n$ as an ``order paramater''. For
  $\alpha < 2$, 
  one has $\Phi = 0$ whereas $\Phi = 1$ for $\alpha > 2$. One can
  however interpolate smoothly between these two limiting cases by
  considering the case $\alpha = 2$ and introducing an additional real
parameter $\mu$ such that
\beq\label{def_mu}
\langle a_k^2 \rangle = e^{-\mu k^2} \;.
\eeq
Performing the same algebra as explained in the Appendix, one obtains the same formula as given in Eq. (\ref{start_expr_rho}) with $u^*(x) = \mu^{-1} \log{x}$. The new variable is thus here $Y = \mu^{-1} \log{x}$ and, setting $Y = \lfloor Y \rfloor + b$ it is easy to see that the
density $\hat \rho_n(Y)$ is given by for $1 \ll Y < n$ 
\beq\label{start_expr_rho_mu}
\hspace*{-2cm}\hat \rho_n(Y) = \frac{\mu}{\pi} \left[  \frac{\sum_{m=-\infty}^\infty (m-b)^2 e^{-\mu(m-b)^2} }{\sum_{m=-\infty}^\infty e^{-\mu(m-b)^2}} - \left[ 
\frac{\sum_{m=-\infty}^\infty (m-b) e^{-\mu(m-b)^2} }{\sum_{m=-\infty}^\infty e^{-\mu(m-b)^2}}\right]^2              \right]^{1/2} \;,
\eeq
which is thus 1-periodic for all $\mu$. In Fig. \ref{fig2} c), one
shows a sketch of $\hat \rho(Y)$ for $\alpha = 2$ given by
Eq. (\ref{start_expr_rho_mu}). For $\mu \to 0$, the
density is almost constant and $\hat \rho_n(Y) \sim
\pi^{-1}\sqrt{\mu/2}$ and the modulation of the density increases
with $\mu$. For large $\mu$, the sum in Eq.~(\ref{start_expr_rho_mu})
is dominated by the terms corresponding to $m=0$ and $m=1$ and $\hat
\rho_n(Y)$ is thus given by a formula similar to Eq. (\ref{pseudo_periodic})
setting $\alpha=2$ and replacing $Y^{\alpha-2}$ by $\mu$. For the
average number of real roots one has 
\begin{eqnarray}
\langle N_n \rangle \propto 
\cases{
\frac{\sqrt{2\mu}}{\pi} n \;,\; \mu \ll 1 \\
n \;,\; \mu \gg 1\;,}
\end{eqnarray}
which shows that this family of real random polynomials (\ref{def_mu})
interpolate smoothly between the cases $\alpha < 2$
(\ref{N_alleq2}) and $\alpha > 2$ (\ref{condensation}).  

\section{A qualitative argument for the transition at $\alpha=2$}

This condensation of the roots on the real axis can be qualitatively
understood if one considers the random polynomials (for $x >0$) $\hat P_n(Y) =
P_n(x)$ of the variable
$Y$, which one writes as 
\bea\label{P_hat}
\hat P_n(Y) = \sum_{k=0}^n \hat a_k w(k,Y) \;, \; w(k,Y) =
\exp{\left[-\frac{1}{2}(k^\alpha - \alpha k Y^{\alpha-1})\right]} \;,
\eea
and $\hat a_k$ are i.i.d. Gaussian variables of unit variance. It is
easy to see that the weights $w(k,Y)$, as a function of $k$, have a
single maximum for $k = Y$ where the second derivative is proportional to
$k^{\alpha - 2}$. Thus for $\alpha > 2$, the weights get more and
more peaked around this maximum for large $k$, whereas $\hat a_k$ is typically of order ${\cal O}(1)$.  
Therefore, given a large integer $m$, $\hat P_n(m)$ is, for $\alpha > 2$, 
dominated by a single term corresponding to $k=m$. Consequently, the sign
of $\hat P_n(m)$ is essentially the sign of $\hat a_m$. This in turn implies that, if $\hat
a_m$ and $\hat a_{m+1}$ have an opposite 
sign, $P_n(x)$ has, with a probability close to $1$, a root in the interval
$[e^{\frac{\alpha}{2}m^{\alpha-1}},e^{\frac{\alpha}{2}(m+1)^{\alpha-1}}]$.
In the case where $\hat a_m$ and $\hat a_{m+1}$ have the same sign, the same
argument shows that $P_n(x)$ has, with a probability close to $1$, a root in the interval
$[-e^{\frac{\alpha}{2}(m+1)^{\alpha-1}},-e^{\frac{\alpha}{2}(m)^{\alpha-1}}]$.
One thus recovers qualitatively the result we had found from the
computation of $\hat \rho_n(Y)$ in Eq. (\ref{condensation}) where we
have shown that $P_n(x)$ has, on average, one root in the interval
$[-e^{\frac{\alpha}{2}(m+1)^{\alpha-1}},-e^{\frac{\alpha}{2}(m)^{\alpha-1}}]
\cup
    [e^{\frac{\alpha}{2}m^{\alpha-1}},e^{\frac{\alpha}{2}(m+1)^{\alpha-1}}]$.   This shows finally that $P_n(x)$ has, on average, $\langle N_n \rangle \propto n$real 
roots. 

We also point out that our argument explains in a
rather intuitive way the result obtained 
by Littlewood and Offord \cite{littlewood} for the random polynomials
$L_n(x)$ (\ref{little}). For these specific polynomials, 
defining $x_0 = 0$, $x_{m} = m^m m !$, they rigorously proved, using
algebraic (and rather cumbersome) methods, that $L_n(x)$ has a root either on $[x_m,x_{m+1}]$
if $\epsilon_m \epsilon_{m+1} =-1$ or in $[-x_{m+1},-x_{m}]$ if
$\epsilon_{m} \epsilon_{m+1} = 1$. Our argument gives some insight on their intriguing result and allows to understand it in a rather simple way.

\section{Conclusion}

To conclude we have introduced a new family of random polynomials
(\ref{def_poly}), indexed by a real $\alpha$. For these random
polynomials, we have computed the mean density of real roots
$\rho_n(x)$ from which we computed the mean number of real roots
$\langle N_n \rangle$
for large $n$. We have shown that, while for $0 \leq \alpha < 1$,
$\langle N_n \rangle \sim (\frac{2}{\pi}) \log{n}$, the behavior of
$\langle N_n \rangle$ for $\alpha > 1$ deviates 
significantly from the logarithmic behavior characteristic for 
Kac's polynomials. For $1< \alpha < 2$, we have shown that $\langle
N_n \rangle
\propto n^{\alpha/2}$ whereas for $\alpha > 2$, $\langle N_n \rangle
\sim n$. This 
family of real random polynomials thus displays an interesting
condensation phenomenon 
of their roots on the real axis, which is accompanied by an ordering
of the roots in 
a quasi periodic structure : this is depicted in Fig. \ref{fig2}. 

Of course, the occurrence of this transition raises several interesting
questions like the behavior of the variance of the number of real
roots for large $n$ as $\alpha$ is varied. It would be also interesting to
compute the two-point correlation function of the 
real roots, which is a rather natural tool to characterize this periodic
structure we have found. In view of this, we hope that this interesting
phenomenon will stimulate further research on random polynomials.

\begin{appendix}
\section{A useful expression for the mean density $\rho_n(x)$}

In this appendix, we derive the expression for the mean density $\rho_n(x)$ as given in Eq. (\ref{start_expr_rho}) 
starting from Eq. (\ref{ek_formula}). We first write $c_n(x) = \langle P_n(x) P_n(x)\rangle$ as
\begin{eqnarray}\label{c_app1}
c_n(x) = e^{-\phi(u^*(x),x)} \sum_{k=0}^n e^{-\tilde \phi(k,x)} \;,
\end{eqnarray}
where $u^*(x)$ is the location of the minimum of $\phi(u,x)$ given in
Eq. (\ref{def_ustar}) 
\beq\label{def_ustar_app}
u^*(x) = \left(\frac{2}{\alpha} \log{x}\right)^{\frac{1}{\alpha-1}} \;,
\eeq
and
\bea\label{phi_tilde}
&& \phi(u^*(x),x) = (1-\alpha)u^*(x)^\alpha  \\
&&\tilde \phi(k,x) = \phi(k,x) - \phi(u^*(x),x) = k^\alpha - \alpha k
	  [u^*(x)]^{\alpha-1} + (\alpha-1) [u^*(x)]^\alpha \;. \nonumber
\eea
The correlator $C_n(x,y) = c_n(\sqrt{xy})$ is given by
Eq. (\ref{c_app1}) together with Eq. (\ref{phi_tilde}) where $x$
is replaced by $\sqrt{xy}$. All the dependence of $C_n(x,y)$ in
$x,y$ is thus contained in $u^*(\sqrt{xy})$ only. From its definition
in Eq.~(\ref{def_ustar_app}) one has immediately
\beq
\partial_x u^*(\sqrt{xy}) = \frac{1}{\alpha(\alpha-1)} \frac{1}{x}
	[u^*(\sqrt{xy})]^{2-\alpha} \;,
\eeq
from which we obtain a set of useful relations
\bea\label{relations}
&&\partial_{x,y}^2 \phi(u^*(\sqrt{xy}),\sqrt{xy}) = -\frac{1}{\alpha(\alpha-1)} \frac{1}{xy} [u^*(\sqrt{xy})]^{2-\alpha}\\
&& \partial_x \tilde \phi(k,\sqrt{xy}) = \frac{1}{x}
(u^*(\sqrt{xy})-k) \nonumber \\
&& \partial_{x,y}^2 \tilde \phi(k,\sqrt{xy}) =
\frac{1}{\alpha(\alpha-1)} \frac{1}{xy} [u^*(\sqrt{xy})]^{\alpha-2}
\;. \nonumber
\eea
For the computation of $\rho_n(x)$ from Eq. (\ref{ek_formula}), it is
useful to introduce the notation, for any function $g(k)$ 
\beq
\langle g(k) \rangle_Z = \frac{\sum_{k=0}^n g(k)\exp{(-\tilde
    \phi(k,\sqrt{xy}))} 
}{\sum_{k=0}^n \exp{(-\tilde \phi(k,\sqrt{xy}))}} \;.
\eeq
From $C_n(x,y) = c_n(\sqrt{xy})$ and
$c_n(x)$ given in Eq.(\ref{c_app1}) one obtains 
\bea\label{last_eq_app}
&&\partial_x \partial_y \log{C_n(\sqrt{xy})} = - \partial_{x,y}^2
\phi(u^*(\sqrt{xy}),\sqrt{xy}) - \langle \partial_x \tilde
\phi(k,\sqrt{xy}) \partial_y \tilde \phi(k,\sqrt{xy})  \rangle_Z
\nonumber \\
&&-
\langle \partial_{x}\tilde \phi(k,\sqrt{xy})  \rangle_Z \langle
\partial_{x}\tilde \phi(k,\sqrt{xy})  \rangle_Z 
- \langle \partial^2_{x,y}\tilde \phi(k,\sqrt{xy})  \rangle_Z \;.
\eea
From the above relations in Eq. (\ref{relations}), it is readily seen
that the first and the last term in Eq. (\ref{last_eq_app}) cancel
each other. Using the relation in Eq. (\ref{ek_formula}), one finally obtains
the relation given in the text in Eq. (\ref{start_expr_rho}).

\end{appendix}

\end{document}